\documentclass[12pt]{revtex4}
\usepackage{graphicx}
\usepackage[english]{babel}
\usepackage{color}

\begin{document}
\title{Radiation Hardness of Graphene and MoS$_2$ Field Effect Devices Against Swift Heavy Ion Irradiation}
\author{$^{1}$O.~Ochedowski, $^{1}$K.~Marinov, $^{2}$G.~Wilbs, $^{2}$G.~Keller, $^{4}$N.~Scheuschner,$^{3}$D.~Severin, $^{3}$M.~Bender,
$^{4}$J.~Maultzsch, $^{2}$F.J.~Tegude, $^{1}$M.~Schleberger\footnote{electronic address: marika.schleberger@uni-due.de}}
\affiliation{$^{1}$Fakult\"at f\"ur Physik and CeNIDE, Universit\"at Duisburg-Essen, Lotharstr.~1, 47048 Duisburg, Germany}
\affiliation{$^{2}$Solid State Electronics Department and CeNIDE, University of Duisburg-Essen, Lotharstr.~55, 47048 Duisburg, Germany}
\affiliation{$^{3}$GSI Helmholtz Centre, Planckstr.~1, 64291 Darmstadt, Germany}
\affiliation{$^{4}$Institut f\"ur Festk\"orperphysik, Technische Universit\"at Berlin, Hardenbergstr.~36, 10623 Berlin, Germany}

\begin{abstract}
We have investigated the deterioration of field effect transistors based on two-dimensional materials due to irradiation with swift heavy ions. Devices were prepared with exfoliated single layers of MoS$_2$ and graphene, respectively. They were characterized before and after irradiation with 1.14~GeV U$^{28+}$ ions using three different fluences. By electrical characterization, atomic force microscopy and Raman spectroscopy we show that the irradiation leads to significant changes of structural and electrical properties. At the highest fluence of $4\times10^{11}$~ions/cm$^2$, the MoS$_2$ transistor is destroyed, while the graphene based device remains operational, albeit with an inferior performance. \end{abstract}
\maketitle
\section{Introduction}
Immediately after its discovery graphene was shown to be an excellent material for the fabrication of field effect devices \cite{Novoselov.2005}. Due to its ballistic transport properties very large mobilities can be achieved at least under optimum conditions \cite{Bolotin.2008,Morozov.2008}. A graphene FET operating at GHz frequencies was reported by IBM researchers in 2008 \cite{Lin.2009}. Because graphene is a gapless semiconductor, it shows an ambipolar behaviour, i.e.~both charge carrier types contribute to its conductivity. Therefore, researchers focussed on other possible 2D materials which have a bandgap and in 2011 the first high performance field effect device with single layer MoS$_2$ was realized \cite{Radisavljevic.2011}. This two-dimensional (2D) material has a lower carrier mobility but also bandgap of $\approx$1.8~eV \cite{Mak.2010} and thus a MoS$_2$-FET can be operated with large off-on current ratios \cite{Yoon.2011}. Both materials are thus promising candidates for use in future electronics \cite{Schwierz.2010,Kim.2011,Das.2013,Bertolazzi.2013}. In this paper we study whether such devices are sensitive to radiation by swift heavy ions (SHI). The reason for this study is two-fold: SHI are a well-known tool for material modification \cite{Avashti.2011,Aumayr.2011} and might thus be useful to manipulate 2D-FETs as well. In addition, this type of projectile interacts with solids primarily via electronic excitations and offer thus the unique chance to study how 2D materials react to ionizing particle irradiation. This is not only interesting for basic science but represents also an important issue if 2D-FETs are to be operational in ionizing environments as e.g.~in outer space.

\section{Experiment}
\subsection{Experimental details}
\begin{figure}
	\centering
	\includegraphics[width=0.7\columnwidth]{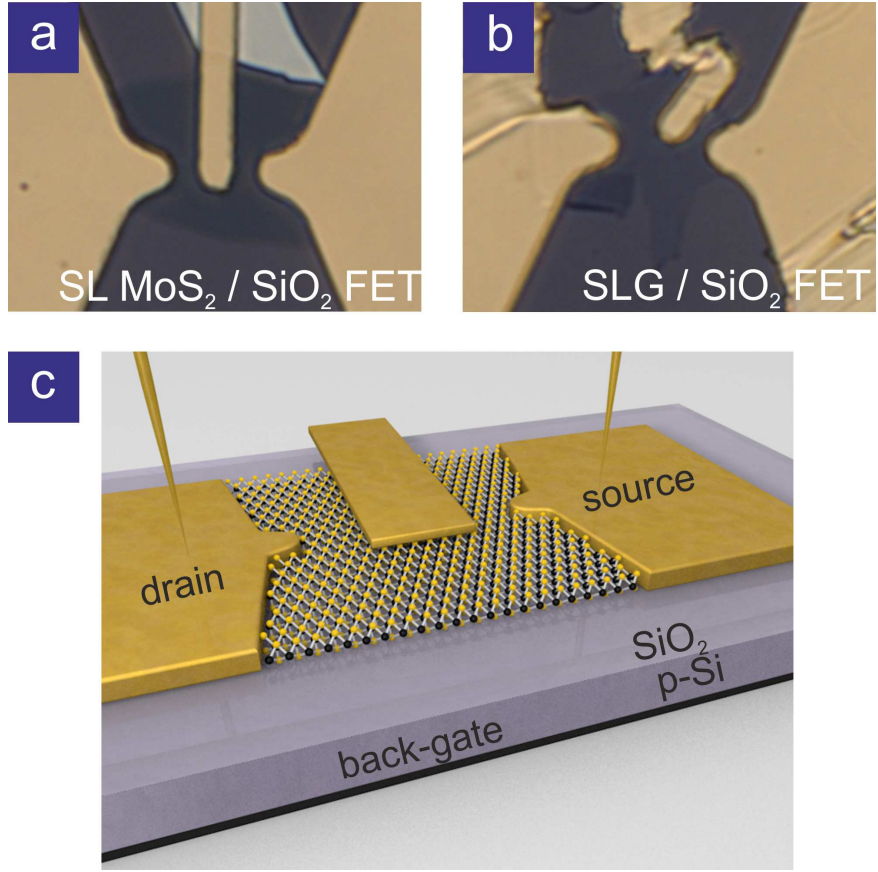}
	\caption{Optical microscopy images of field effect devices based on (a) single layer MoS$_2$ and (b) single layer graphene. The electrical setup is schematically shown in (c). The third contact is not used and was kept floating during measurements.}
	\label{FETs}
\end{figure}

For this experiment simple field effect devices (called FETs in the following) were prepared from single layer graphene (SLG) as well as from single layer MoS$_2$ (SLM). The first step is mechanical exfoliation from single crystals onto 90~nm SiO$_2$ on a Si wafer ($p$-doped, $\rho=(0.001-0.005)~\Omega$cm). The second step is selection of appropriate flakes and determination of their layer number with $\mu$-Raman spectroscopy \cite{Ferrari.2006}. In a third step selected SLG and SLM are contacted using photolithography and vacuum evaporation. Two gold contacts (with Ti as bonding agent) serve as source and drain, the Si substrate as a global backgate, see fig~\ref{FETs}. 

Typical channel length and width of our devices are $L=6~\mu$m and $W=(3-12)~\mu$m. Step four is the electrical characterization of the FETs before irradiation with respect to their output ($I_D(U_{DS})$) and transfer ($I_D(U_{GS})$) characteristics. Important physical properties as conductivity $\sigma$ and mobility $\mu$ are directly derived from the experiment as follows:
$$\sigma=dI_D/dU_{DS},~~\mathrm{and}~\mu=\frac{dI_{D}}{dU_{GS}}\cdot\frac{L}{WC_iU_{DS}}.$$ Here, $I_D$ is the drain current, $U_{GS}$ is the voltage between gate and source, $U_{DS}$ is the voltage between drain and source, see fig.~\ref{FETs}. $C_i=\frac{\epsilon\epsilon_r}{d}=3.837 \cdot 10^{-4}$~F/m$^2$ is the capacitance between the channel and the backgate per unit area, $\epsilon$ and $\epsilon_r$ are the dielectric constants of air and of the dielectric (SiO${_2}$:~$\epsilon_r=3.9$ ), respectively, and $d$ is the thickness of the oxide layer (90~nm). In addition, the charge carrier density can be directly derived from this data by: $$~n_{e,h}=\frac{\sigma}{e\mu_n}$$ with $e$ the elementary charge. Note, that $\mu$ and $\sigma(U_{GS}=0\mathrm~{V})$ are measured independently from each other, while $n(U_{GS}=0\mathrm~{V})$ is calculated from $\sigma$ and $\mu$.

The FET characteristics that we find are typical and comparable to literature data \cite{Venugopal.2011,Late.2012}, however device performance is definitively below the reported record values, see below. In general we found a great device-to-device variation with respect to transistor parameters (see figs.~\ref{FETMoS2} and \ref{FETSLG}). In addition, we observed that MoS$_2$ transistors show significant aging even if stored under N$_2$ atmosphere \cite{Late.2012}, while graphene devices appear to be stable for weeks. Therefore, SLM FETs for this study have been prepared immediately before irradiation to avoid contamination problems. 

In the fifth step the samples were irradiated with 1.14~GeV $^{238}$U in the UHV irradiation set-up at the M-branch of the swift heavy ion accelerator at the GSI in Darmstadt, Germany. For each type of FET three different fluences $\varphi$ were chosen: $4\times10^{10},~1.5\times10^{11}~\mathrm{and}~ 4\times10^{11}$ ions/cm$^2$ (corresponding to 400, 1500 and 4000~ions/$\mu$m$^2$). In bulk materials the radius of a typical SHI ion track (i.e.~the permanently modified region) is about one nm. This means that at the chosen fluences only the last fluence definitely results in overlapping tracks while the lowest one grants individual impacts. The projected range of the uranium ions at this energy is $\approx$~46 microns \cite{Ziegler.2010}. It is therefore safe to assume that they completely pass the SiO$_2$ layer. The uranium ions themselves have thus no further influence on the electronic properties of the devices and any changes are consequences of the processes triggered by the projectiles.

For comparison we irradiated additional SLM and SLG samples without electrical contacts with $4\times10^{11}$~ions/cm$^2$ (high fluence) and under otherwise identical conditions chosen for the FETs. These samples were analyzed with atomic force microscopy (AFM, VECOO Dimension-3100; Nanosensors NCHR tips) in tapping mode and in the case of SLG also with Micro-Raman spectroscopy (LabRAM HR, Horiba Jobin Yvon), $E_{L}=$2.33~eV, 1.96~eV and 1.49~eV, $P_{Laser}\leq1$~mW). 

\subsection{Results}
\begin{figure}
	\centering
	\includegraphics[width=\columnwidth]{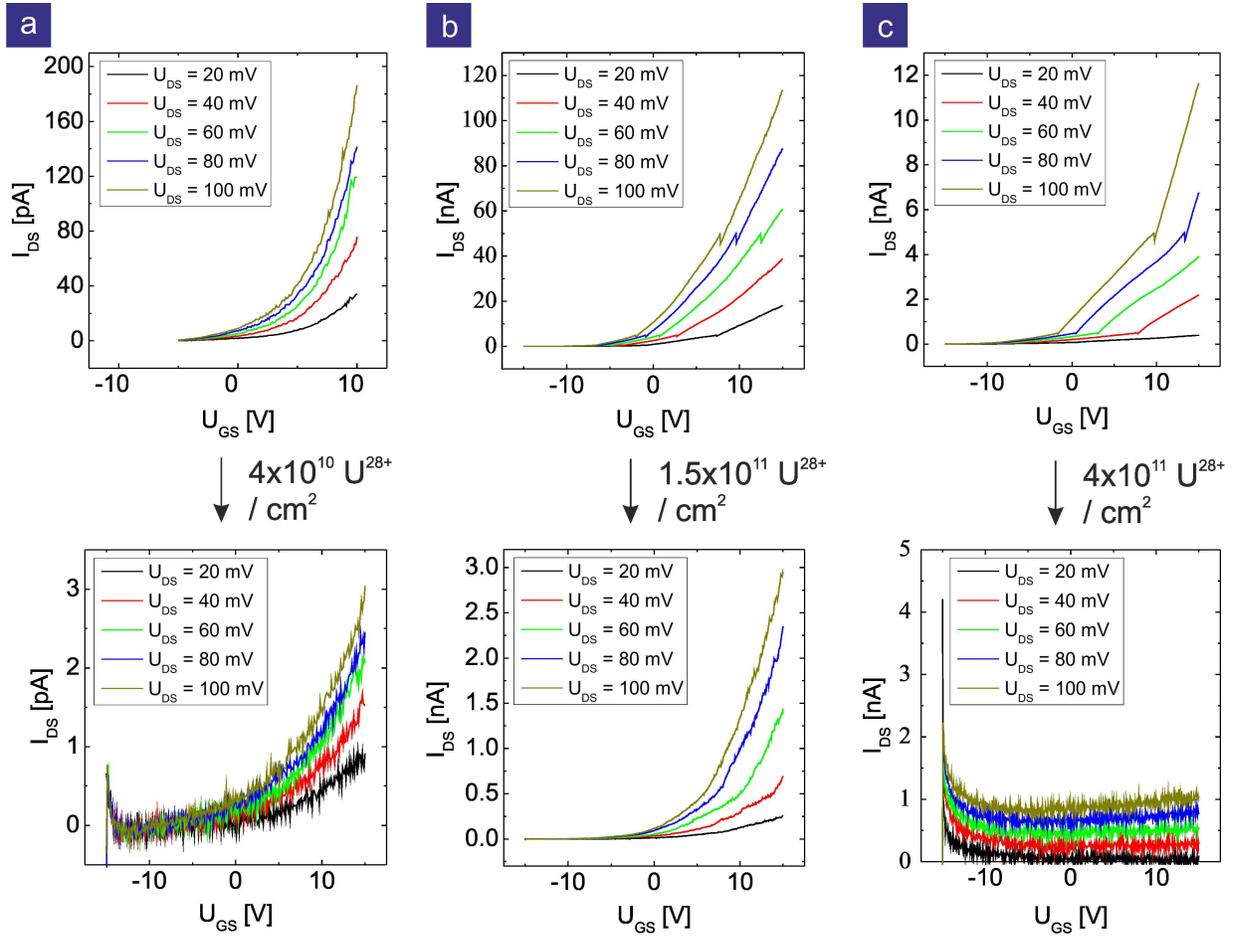}
	\caption{Transfer characteristics of SLM FET devices. Upper panel shows $I_D(U_{GS})$ before irradiation for the three different devices. Lower panel shows $I_D(U_{GS})$ as recorded after irradiation with $^{238}$U ions at a fluence of (a) $4\times10^{10}$ ions/cm$^2$, (b) $1.5\times10^{11}$ ions/cm$^2$ and (c) $4\times10^{11}$ ions/cm$^2$, respectively.}
	\label{FETMoS2}
\end{figure}

First, we present the data for SLM FETs. The results for all three devices are shown in fig.~\ref{FETMoS2}, where we have plotted the transfer characteristics, i.e.~the drain current as a function of the gate voltage. Before irradiation (upper panel) the SLM FETs show basically the same characteristics as reported in literature. In our devices amplification sets in at a gate voltage of about $U_{GS}=-4$~V (pinch-off). Typical values for mobility are $(2.5\cdot10^{-4}-1.7\cdot10^{-1})$~cm$^2$/Vs and for charge carrier concentration $(9.8\cdot10^{11}-2.5\cdot10^{13})~n_e$/cm$^2$. 

After irradiation (fig.~\ref{FETMoS2}, lower panel) those quantities show significant changes. At low fluences (a) we observe a decreased drain current by almost two orders of magnitude. Also for the intermediate fluence (b) the conductivity of the SLM FETs deteriorates. The corresponding drain current has decreased by $\approx$~1.5 orders of magnitude with respect to the pristine SLM FET. Note however, that the device (b) was of higher quality to begin with, compared to the SLM FET exposed to the low fluence irradiation (a): The latter controls a current about three orders of magnitude less, see fig.~\ref{FETMoS2}. For the highest fluence (c), the SLM FET is no longer operational. To verify this result another device of this type was subjected to a high fluence irradiation. Also this second SLM FET was rendered non-functional by the irradiation.

\begin{figure}
	\centering
	\includegraphics[width=\columnwidth]{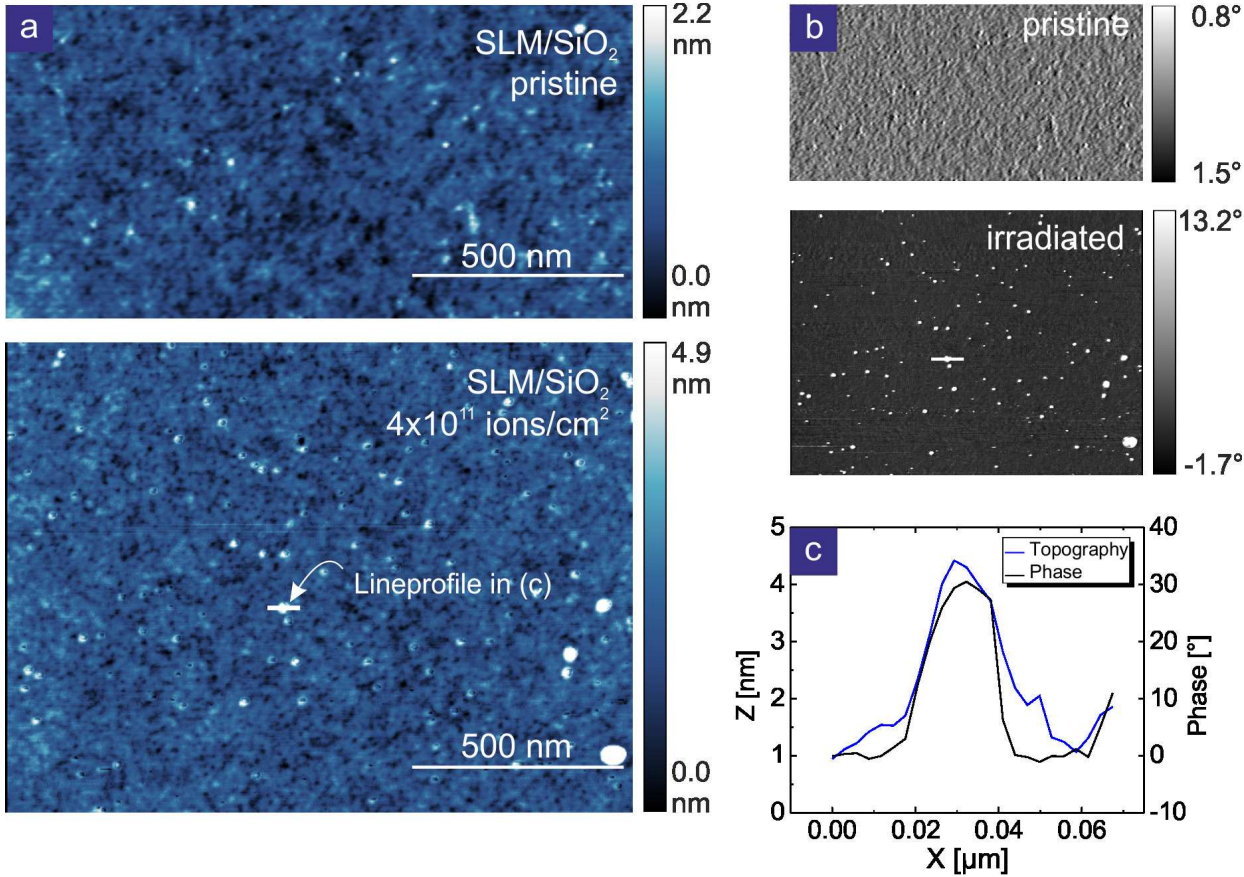}
	\caption{(a) AFM topography image of SLM before (upper panel) and after irradiation (lower panel) with $4\times10^{11}$ U ions/cm$^2$. After irradiation the surface is covered with protrusions and small holes, the former can be clearly identified in the phase image (b). Their height is about a few nm, diameters can be as large as 50~nm.}
	\label{AFMMoS2}
\end{figure}

Atomic force microscopy images of single layer MoS$_2$ exposed to the highest fluence show many randomly distributed hillocks (see figs.~\ref{AFMMoS2}a and b, lower panels) which are not present on pristine samples (upper panels in figs.~\ref{AFMMoS2}a and b). Their apparent height depends strongly on the scanning parameters and they can be seen even more clearly in the corresponding phase images. These protrusions cover about 1.62\% of the surface.

\begin{figure}
	\centering
	\includegraphics[width=\columnwidth]{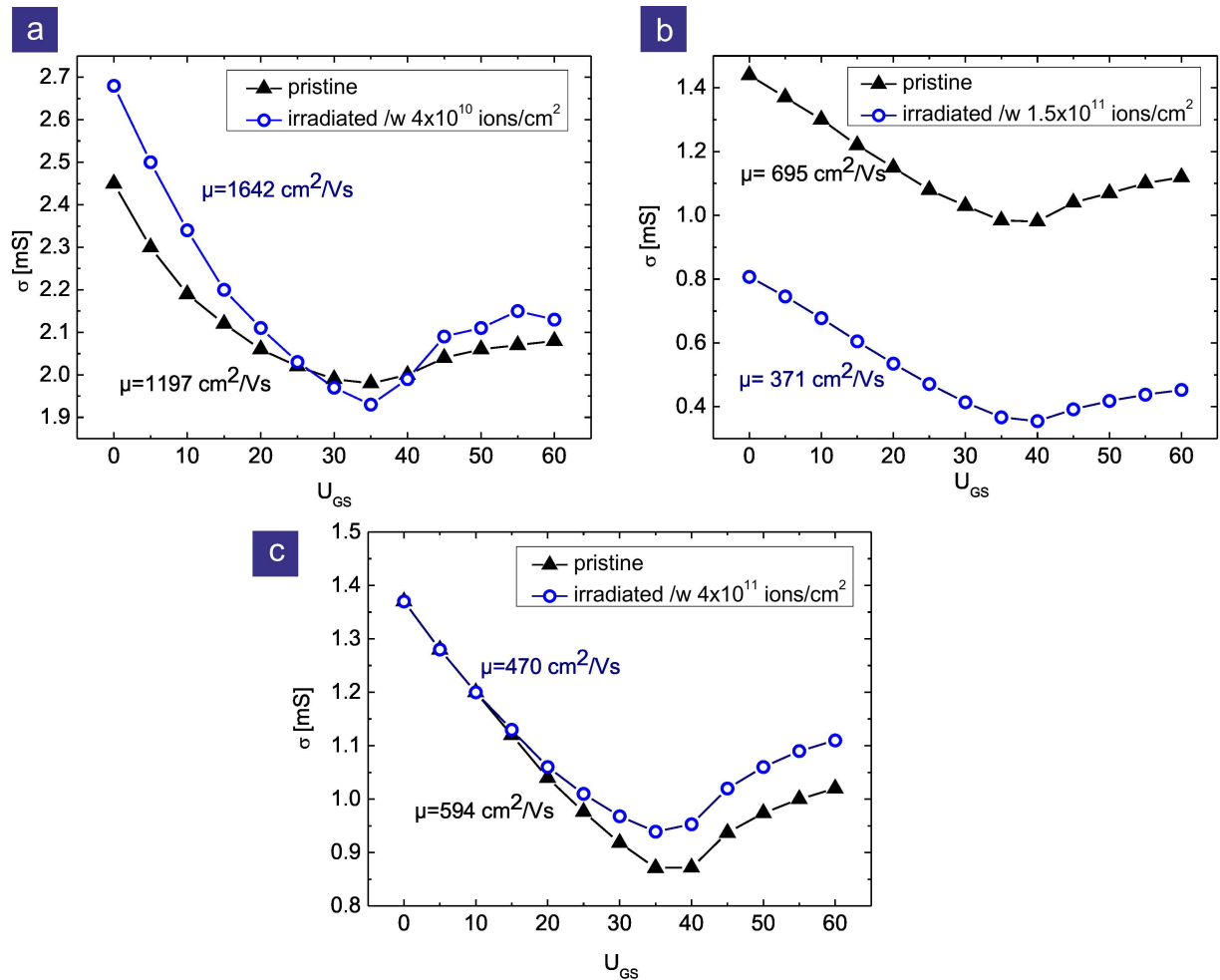}
	\caption{Conductivity of three different SLG-FETs as a function of gate voltage $U_{GS}$ before (black triangles) and after irradiation (blue circles) with $4\times10^{10}$ ions/cm$^2$, (b) $1.5\times10^{11}$ ions/cm$^2$ and (c) $4\times10^{11}$ ions/cm$^2$, respectively.}
	\label{FETSLG}
\end{figure}

Second, we present the data for the SLG FETs. The results for all three devices are shown in fig.~\ref{FETSLG}, where we have plotted the conductivity as a function of the gate voltage. Before irradiation (black triangles) the SLG FETs show the ambipolar behaviour typical for graphene devices. All SLG FETs have excess carriers resulting in a $p$-type doping ranging from $(1.3\cdot10^{13}-1.4\cdot10^{13})$~ions/cm$^2$. Mobility values range from $(243-390)$~cm$^2$/Vs for electrons and for holes we find $(595-1198)$~cm$^2$/Vs. 

After irradiation (fig.~\ref{FETSLG}, blue circles) with a low fluence (a) the mobility of the SLG FET increases significantly for holes as well as electrons while the carrier density slightly decreases. For the intermediate fluence (b) the mobility decreases significantly 
while the carrier density slightly increases. Even after irradiation with the highest fluence (c), the SLG FET is still operational in contrast to the SLM FET. Carrier density has increased, hole mobility slightly decreased upon irradiation. Another SLG device was exposed to high fluence irradiation and remained operational as well.

\begin{figure}
	\centering
	\includegraphics[width=\columnwidth]{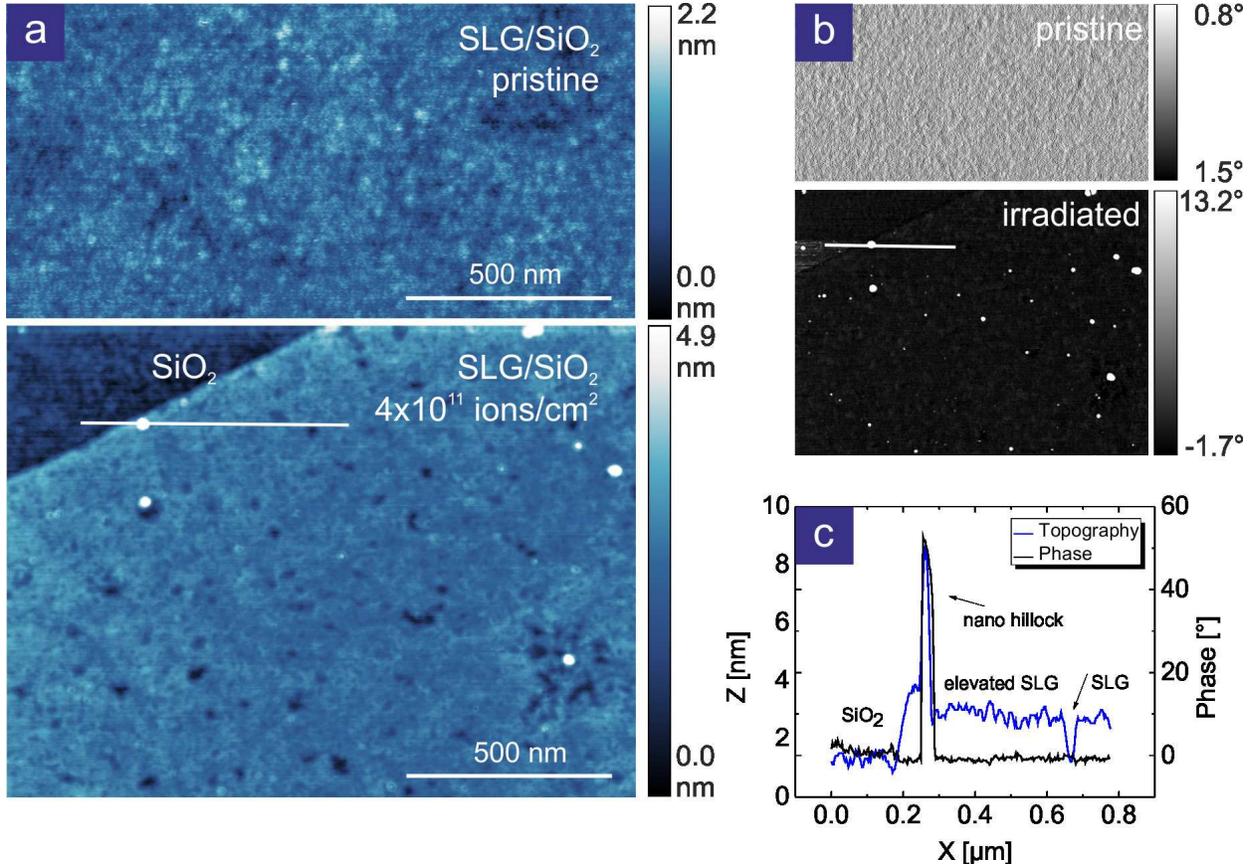}
	\caption{(a) AFM topography image of pristine (upper panel) SLG and after irradiation (lower panel) with $4\times10^{11}$U ions/cm$^2$. After irradiation the surface is covered with protrusions which can be clearly identified in the phase image (b). The hillocks height reaches 10~nm, the holes depth is limited to the apparent thickness of the SLG. The hillock diameter can be as large as 50~nm.}
	\label{AFMSLG}
\end{figure}

Atomic force microscopy images of single layer graphene exposed to the highest fluence show many randomly distributed hillocks and rather large pits (see fig.~\ref{AFMSLG}) which are not present on pristine samples. The overall appearance is similar to what has been found for SLM. The apparent height of the hillocks depends strongly on the scanning parameters. Because they can be seen even more clearly in the corresponding phase images, this data was analyzed to determine that protrusions cover about 1.09\% of the surface. 

\begin{figure}
	\centering
	\includegraphics[width=0.6\columnwidth]{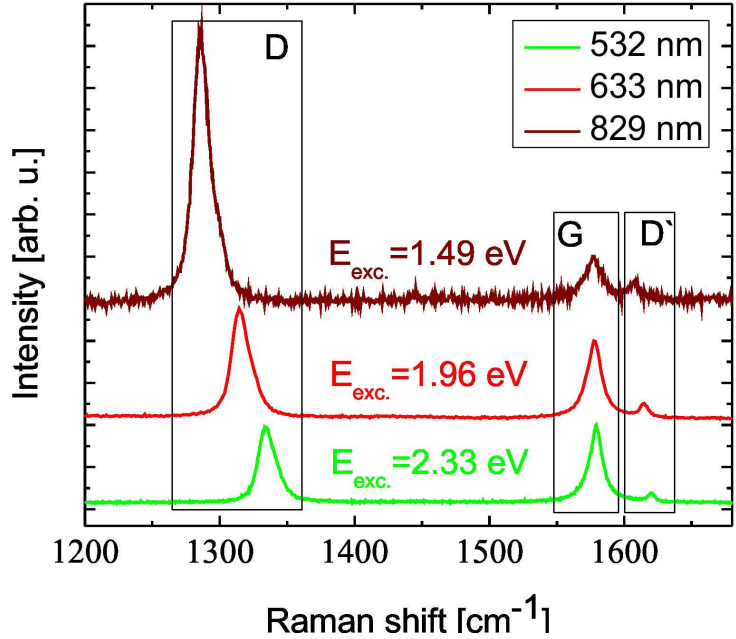}
	\caption{Raman spectra of SLG after irradiation with $4\times10^{11}$U ions/cm$^2$ (high fluence) for three different excitation wave lengths. The $D$ peak indicating irradiation induced defects is clearly present. The normalized $E_L^4(I_D/I_G)$ peak height ratio can be used to estimate the induced damage to the SLG sheet.}
	\label{Raman}
\end{figure}

In order to further analyze the nature of the induced defects in graphene, Raman spectroscopy was performed. This method has been proven to be a powerful tool to characterize defects in SLG \cite{Lucchese.2010,Cancado.2011,Jorio.2010,MartinsFerreira.2010}. The signature of defects is given by the double-resonant $D$ and $D\prime$ peaks at approximately 1350 cm$^{-1}$ and 1620 cm$^{-1}$ \cite{Thomsen.2000,Maultzsch.2004}. In fig.~\ref{Raman}, we show Raman spectra of a similar graphene flake as shown in fig.~\ref{AFMSLG}, irradiated with $4\times10^{11}$~ions/cm$^2$. Spectra have been taken with three different laser excitation energies enabling us to compare our results with literature data from Ar$^+$ irradiated graphene (see discussion).

\subsection{Discussion}
The data above indicates that the irradiation of 2D-FETs with SHI can have adverse effects: It can lead to inferior device performance (e.g.~high fluences SLG FET) up to total destruction (SLM FET) but it can also cause an improved device performance (low fluences SLG FET). This immediately suggests that at least two counteracting mechanisms could play a role here. We propose that two relevant mechanisms are 1) doping and 2) defect creation. The latter seems to be straightforward as ion irradiation is known to cause structural defects, even in graphene \cite{Ahlgren.2011,Lehtinen.2010,Krasheninnikov.2010,Akcoeltekin.2011,Mathew.2011,Xue.2012,Deretzis.2012}. By using a combined scanning electron microscope/focused ion beam system it has been shown, that graphene possesses an enhanced resistance towards sputtering \cite{Lopez.2010}. However, here we have to keep in mind that the cross section for direct collisions of swift heavy ions with target atoms is negligible. In order for this mechanism to be in effect, we have to postulate that the electronic excitation caused by the SHI is the origin of the defect creation. This is in agreement with our finding here that SLM FETs are more easily destroyed than SLG FETs. In the latter material the electronic energy will be spread very efficiently so that the energy density at a given time is too low to create large defects, while in MoS$_2$ (being a semiconductor) the electronic energy dissipation will be less rapid and significant damage will occur more easily. Any damage to the lattice will act as scattering center and will therefore result in reduced mobility. 

The second mechanism, doping, will lead to an increased number of charge carriers and thus give rise to an enhanced conductivity (at a given mobility).
Here, doping could be achieved in various ways: ions may either directly induce interstitials and vacancies or their interaction with the substrate gives rise to modifications of the graphene lattice. In this case, also substitutional doping would be possible in the following way \cite{Ochedowski.2013}. After passing through the 2D material the projectile enters the substrate where energy is deposited along its track. As the dielectric substrate material SiO$_2$ cannot dissipate the energy effectively, the primary electronic excitation is transferred to the lattice in a very small volume surrounding the trajectory. This heated zone is called a {\it thermal spike} \cite{Toulemonde.1992} and causes material to be ejected at the impact sites in the near-surface regions \cite{Akcoltekin.2007}. This foreign material can be caught by graphene \cite{Ochedowski2.2013} and act as donor or acceptor, respectively. Substitutional doping by irradiation thus introduces additional charge carriers which might either increase the total charge carrier density or decrease it, depending on the type of carrier and the initial doping of the 2D material. Note, that whether changes in carrier concentration are due to  doping and/or to the removal of adsorbates cannot be distinguished at this point.  

In the case of SLM FETs, doping by SHI irradiation seems to be only a minor effect and the deterioration of device performance with increasing fluence can be attributed solely to the increasing number of ion induced defects. The AFM images (fig.~\ref{AFMMoS2}) clearly show that the SLM sustains significant structural damage due to ion irradiation. At the highest fluence ion tracks overlap and quite obviously material is ejected from the surface. The protrusions could be the remnants of this process. 

In case of the SLG FETs (see fig.~\ref{FETSLG}) both mechanisms seem to play a role. Irradiation with SHI can increase the number of charge carriers (doping) but at the same time reduce their mean free path (defects). In order to further discuss our results we focus in the following on the primary quantity which determines the performance of the SLG FET, i.e.~the hole mobility. 

Our experiment shows that the performance of an SLG FET with average mobility (see fig.~\ref{FETSLG}a) increases after low fluence irradiation. This would indicate that either scattering centers have been removed and/or electron density has increased by doping. Pristine exfoliated graphene on SiO$_2$ is usually $p$-doped due to adsorbates. The allegedly {\it reduced} hole density determined at $U_{GS}=0$ is fully consistent with e.g.~ substitutional doping, as the introduction of a given number of electrons from foreign atoms would compensate a corresponding number of holes and thus lead to an overall decreased holes density, while the hole conductivity at $U_{GS}=0$ increases. Scattering centers related to doping at this fluence are still far enough apart: assuming point like defects the average distance $l$ between to impacts is proportional to $\sqrt{1/\varphi}$ \cite{MartinsFerreira.2010} and thus amounts to 50~nm . 

We also find that the SLG FET with a below-average performance (see fig.~\ref{FETSLG}c) remains more or less intact even after high fluence irradiation, i.e.~the decrease in hole mobility is probably compensated by the increasing hole density. It is quite surprising that the degree of structural damage sustained by graphene seems to be comparable to that observed in MoS$_2$. Both, AFM as well as Raman data show that the SLG is highly defective after high fluence irradiation, see figs.~\ref{AFMSLG} and \ref{Raman}. Nevertheless, the SLG FETs were still operational, proving the superior resistance of graphene devices to ionizing particle radiation as predicted using atomistic simulations \cite{Ahlgren.2012}. The biggest effect of irradiation is observed if an SLG FET of below-average performance (see fig.~\ref{FETSLG}b) is irradiated with a medium fluence. Here, the decrease in mobility is the largest of all three irradiated SLG FETs and the very slight increase in carrier density can seemingly not compensate this effect. 

Transport characteristics of FETs are governed by a variety of factors such as unintentional channel doping due to substrate interactions, environmental adsorbates, contacts or fabrication steps, as well as oxide thickness, channel width and length \cite{Chen.2009,Han.2011,Joshi.2010,Late.2012,Ganapathi.2013}. In devices where the latter is on the order of the depletion-layer widths of the source and drain junction, the carrier density does not necessarily remain constant at high gate voltages and short-channel effects must be considered \cite{Alarcon.2013}. In order to investigate these effects and to clarify the influence of ion irradiation induced contributions, experiments with dedicated devices are currently underway.

Finally, we discuss our Raman data. With respect to ion irradiation of graphene, it has been shown that point defects can be introduced in the graphene lattice most efficiently by low energy ion bombardment \cite{Lucchese.2010}. In order to estimate the density of defects which were created by SHI irradiation, we determined the amplitude ratio of the defect activated $D$ mode to the $G$ mode ($I_D/I_G$) multiplied by $E_L^4$ (where $E_L$ is the Laser energy) to account for the energy dependence of $I_D/I_G$. Following the method from \cite{Lucchese.2010,Cancado.2011} we find for our sample an $E_L^4(I_D/I_G)$ ratio of $\approx$32 which corresponds to $(7.5\pm2.3)\times10^{11}$~defects/cm$^2$. Because the sample was irradiated with only $4\times$10$^{11}$~ion/cm$^2$, defects induced by SHI seem to exceed single carbon atom displacements caused by 90~eV Ar$^+$ ions. From this we conclude, that the electronic excitation induced by the SHI in the graphene, the substrate or both, is able to create larger defects with a higher Raman cross section in the graphene flake than Ar$^+$ ions. This unexpected effect will be studied in more detail in a forthcoming experiment.

\subsection{Conclusions}
We have shown that 2D FETs show significant changes with respect to output and transfer characteristics after irradiation with swift heavy ions. Devices may improve or deteriorate after irradiation, a finding which could be interpreted in terms of two mechanisms working in opposite directions. Especially our result that single layer graphene FETs can be improved with respect to conductivity might prove very interesting for applications. The  doping mechanism could be used to increase the number of carriers without introducing too many scattering defects. It would be interesting to study this effect in devices which have been optimized before irradiation, e.g.~by heating to remove adsorbates. This would also enable a more detailed study of the superior resistance of graphene FETs to ionizing particle irradation in comparison to FETs based on MoS$_2$.

\subsection{Acknowledgments}
We acknowledge financial support from the BMBF in the framework of the Research project {\em Modifikation von Systemen reduzierter Dimension durch Ionenbeschuss} and from the DFG in the framework of the Priority Program 1459 {\em Graphene} (O.O., N.S.), the SFB 616 {\em Energy dissipation on surfaces} (K.M.) and from the ERC under grant number 259286 (J.M.).

\section{References}

\begin{thebibliography}{10}

\bibitem{Novoselov.2005}
K.~S. Novoselov, A.~K. Geim, S.~V. Morozov, D.~Jiang, M.~I. Katsnelson, I.~V.
  Grigorieva, S.~V. Dubonos, and A.~A. Firsov.
\newblock {\em Nature}, 438(7065):197--200, 2005.

\bibitem{Bolotin.2008}
K.~I. Bolotin, K.~J. Sikes, Z.~Jiang, M.~Klima, G.~Fudenberg, J.~Hone, P.~Kim,
  and H.~L. Stormer.
\newblock {\em Solid State Communications}, 146(9-10):351--355, 2008.

\bibitem{Morozov.2008}
S.~Morozov, K.~Novoselov, M.~Katsnelson, F.~Schedin, D.~Elias, J.~Jaszczak, and
  A.~Geim.
\newblock {\em Physical Review Letters}, 100(1), 2008.

\bibitem{Lin.2009}
Y.-M. Lin, K.~A. Jenkins, A.~Valdes-Garcia, J.~P. Small, D.~B. Farmer, and
  P.~Avouris.
\newblock {\em Nano Letters}, 9(1):422--426, 2009.

\bibitem{Radisavljevic.2011}
B.~Radisavljevic, A.~Radenovic, J.~Brivio, V.~Giacometti, and A.~Kis.
\newblock {\em Nature Nanotechnology}, 6(3):147--150, 2011.

\bibitem{Mak.2010}
K.~F. Mak, C.~Lee, J.~Hone, J.~Shan, and T.~F. Heinz.
\newblock {\em Physical Review Letters}, 105(13), 2010.

\bibitem{Yoon.2011}
Y.~Yoon, K.~Ganapathi, and S.~Salahuddin.
\newblock {\em Nano Letters}, 11(9):3768--3773, 2011.

\bibitem{Schwierz.2010}
F.~Schwierz.
\newblock {\em Nature Nanotechnology}, 5(7):487--496, 2010.

\bibitem{Kim.2011}
K.~Kim, J.-Y. Choi, T.~Kim, S.-H. Cho, and H.-J. Chung.
\newblock {\em Nature}, 479(7373):338--344, 2011.

\bibitem{Das.2013}
S.~Das, H.-Y. Chen, A.~V. Penumatcha, and J.~Appenzeller.
\newblock {\em Nano Letters}, 13(1):100--105, 2013.

\bibitem{Bertolazzi.2013}
S.~Bertolazzi, D.~Krasnozhon, and A.~Kis.
\newblock {\em ACS Nano}, page 130319125345005, 2013.

\bibitem{Aumayr.2011}
F.~Aumayr, S.~Facsko, A.~S. El-Said, C.~Trautmann, and M.~Schleberger.
\newblock {\em Journal of Physics: Condensed Matter}, 23(39):393001, 2011.

\bibitem{Avashti.2011}
Eds:~D.K. Avashti and G.K. Mehta.
\newblock Swift heavy ions for materials engineering and nanostructuring.
\newblock {\em Springer Series In Materials Science}, 2011.

\bibitem{Ferrari.2006}
A.~C. Ferrari, J.~C. Meyer, V.~Scardaci, C.~Casiraghi, M.~Lazzeri, F.~Mauri,
  S.~Piscanec, D.~Jiang, K.~S. Novoselov, S.~Roth, and A.~K. Geim.
\newblock {\em Physical Review Letters}, 97(18), 2006.

\bibitem{Venugopal.2011}
A.~Venugopal, J.~Chan, X.~Li, C.~W. Magnuson, W.~P. Kirk, L.~Colombo, R.~S.
  Ruoff, and E.~M. Vogel.
\newblock {\em Journal of Applied Physics}, 109(10):104511, 2011.

\bibitem{Late.2012}
D.~J. Late, B.~Liu, H.~S. S.~R. Matte, V.~P. Dravid, and C.~N.~R. Rao.
\newblock {\em ACS Nano}, 6(6):5635--5641, 2012.

\bibitem{Ziegler.2010}
J.~F. Ziegler, M.~D. Ziegler, and J.~P. Biersack.
\newblock {\em Nuclear Instruments and Methods in Physics Research Section B:
  Beam Interactions with Materials and Atoms}, 268(11-12):1818--1823, 2010.

\bibitem{Lucchese.2010}
M.~M Lucchese, F.~Stavale, E.~H~M. Ferreira, C.~Vilani, M.~V.~O. Moutinho,
  R.~B. Capaz, C.~A. Achete, and A.~Jorio.
\newblock {\em Carbon}, 48(5):1592--1597, 2010.

\bibitem{Cancado.2011}
L.~G. Cancado, A.~Jorio, E.~H.~M. Ferreira, F.~Stavale, C.~A. Achete, R.~B.
  Capaz, M.~V.~O. Moutinho, A.~Lombardo, T.~S. Kulmala, and A.~C. Ferrari.
\newblock {\em Nano Letters}, 11(8):3190--3196, 2011.

\bibitem{Jorio.2010}
A.~Jorio, M.~M. Lucchese, F.~Stavale, E.~H. Martins~Ferreira, M.~V.~O.
  Moutinho, R.~B. Capaz, and C.~A. Achete.
\newblock {\em Journal of Physics: Condensed Matter}, 22(33):334204, 2010.

\bibitem{MartinsFerreira.2010}
E.~H. Martins~Ferreira, M.~V.~O. Moutinho, F.~Stavale, M.~M. Lucchese, R:~B.
  Capaz, C.~A. Achete, and A.~Jorio.
\newblock {\em Physical Review B}, 82(12), 2010.

\bibitem{Thomsen.2000}
C.~Thomsen and S.~Reich.
\newblock {\em Physical Review Letters}, 85(24):5214--5217, 2000.

\bibitem{Maultzsch.2004}
J.~Maultzsch, S.~Reich, and C.~Thomsen.
\newblock {\em Physical Review B}, 70(15), 2004.

\bibitem{Ahlgren.2011}
E.~H. {\AA}hlgren, J.~Kotakoski, and A.~V. Krasheninnikov.
\newblock {\em Physical Review B}, 83(11), 2011.

\bibitem{Lehtinen.2010}
O.~Lehtinen, J.~Kotakoski, A.~V. Krasheninnikov, A.~Tolvanen, K.~Nordlund, and
  J.~Keinonen.
\newblock {\em Physical Review B}, 81(15), 2010.

\bibitem{Krasheninnikov.2010}
A.~V. Krasheninnikov and K.~Nordlund.
\newblock {\em Journal of Applied Physics}, 107(7):071301, 2010.

\bibitem{Mathew.2011}
S.~Mathew, T.~K. Chan, D.~Zhan, K.~Gopinadhan, A.~Roy~Barman, M.~B.~H. Breese,
  S.~Dhar, Z.~X. Shen, T.~Venkatesan, and J.~T.~L. Thong.
\newblock {\em Journal of Applied Physics}, 110(8):084309, 2011.

\bibitem{Deretzis.2012}
I.~Deretzis, G.~Piccitto, and A.~La~Magna.
\newblock {\em Nuclear Instruments and Methods in Physics Research Section B:
  Beam Interactions with Materials and Atoms}, 282:108--111, 2012.

\bibitem{Xue.2012}
S.~Zhao and J.~Xue and Y.~Wang and S.~Yan.
\newblock {\em Nanotechnology}, 23:285703, 201.

\bibitem{Akcoeltekin.2011}
S.~Akc\"oltekin, H.~Bukowska, T.~Peters, O.~Osmani, I.~Monnet, I.~Alzaher,
  B.~Ban~d'Etat, H.~Lebius, and M.~Schleberger.
\newblock {\em Applied Physics Letters}, 98(10):103103, 2011.

\bibitem{Lopez.2010}
J.~J. Lopez, F.~Greer, and J.~R. Greer.
\newblock {\em Journal of Applied Physics}, 107(10):104326, 2010.

\bibitem{Ochedowski.2013}
O.~Ochedowski, B.~Kleine~Bussmann, B.~Ban~d´Etat, H.~Lebius, and
  M.~Schleberger.
\newblock {\em Applied Physics Letters}, in press, 2013.

\bibitem{Toulemonde.1992}
M.~Toulemonde, C.~Dufour, and E.~Paumier.
\newblock {\em Physical Review B}, 46(22):14362--14369, 1992.

\bibitem{Akcoltekin.2007}
E.~Akc{\"o}ltekin, T.~Peters, R.~Meyer, A.~Duvenbeck, M.~Klusmann, I.~Monnet,
  H.~Lebius, and M.~Schleberger.
\newblock {\em Nature Nanotechnology}, 2(5):290--294, 2007.

\bibitem{Ochedowski2.2013}
O.~Ochedowski, S.~Akc\"oltekin, B.~Ban~d´Etat, H.~Lebius, and M.~Schleberger.
\newblock {\em Nuclear Instruments and Methods in Physics Research Section B:
  Beam Interactions with Materials and Atoms}, accepted, 2013.

\bibitem{Ahlgren.2012}
E.~H. {\AA}hlgren, J.~Kotakoski, O.~Lehtinen, and A.~V. Krasheninnikov.
\newblock {\em Applied Physics Letters}, 100(23):233108, 2012.

\bibitem{Chen.2009}
F.~Chen, J.~Xia, D.~K. Ferry, and N.~Tao.
\newblock {\em Nano Letters}, 9(7):2571--2574, 2009.

\bibitem{Han.2011}
S.-J. Han, Z.~Chen, A.~A. Bol, and Y.~Sun.
\newblock {\em IEEE Electron Device Letters}, 32(6):812--814, 2011.

\bibitem{Joshi.2010}
P.~Joshi, H.~E. Romero, A.~T. Neal, V.~K. Toutam, and S.~A. Tadigadapa.
\newblock {\em Journal of Physics: Condensed Matter}, 22(33):334214, 2010.

\bibitem{Ganapathi.2013}
K.~Ganapathi, Y.~Yoon, M.~Lundstrom, and S.~Salahuddin.
\newblock {\em IEEE Transactions on Electron Devices}, 60(3):958--964, 2013.

\bibitem{Alarcon.2013}
A.~Alarcon, V.-H. Nguyen, S.~Berrada, D.~Querlioz, J.~Saint-Martin, A.~Bournel,
  and P.~Dollfus.
\newblock {\em IEEE Transactions on Electron Devices}, 60(3):985--991, 2013.

\end{thebibliography}

\end{document}